\documentclass[epj]{svjour}
\usepackage{graphicx}

\let\I\i
\def\i{\mathrm{i}}
\def\e{\mathrm{e}}
\def\d{\mathrm{d}}
\def\half{{\textstyle{1\over2}}}

\def\h{{\scriptscriptstyle{1\over2}}}
\def\th{{\scriptscriptstyle{3\over2}}}

\def\vec#1{\mbox{\boldmath$#1$}}

\def\CG#1#2#3#4#5#6{C^{#5#6}_{#1#2#3#4}}

\begin{document}

\title{A chiral quark model for meson electro-production 
in the S11 partial wave}

\author{%
B.~Golli\inst{1}
\and
S.~\v{S}irca\inst{2}}

\institute{%
Faculty of Education,
              University of Ljubljana and J.~Stefan Institute,
              1000 Ljubljana, Slovenia
\and
Faculty of Mathematics and Physics,
              University of Ljubljana and J.~Stefan Institute,
              1000 Ljubljana, Slovenia}

\date{\today}

\abstract{%
We calculate the meson scattering and electroproduction amplitudes in 
the S11 partial wave in a coupled-channel approach that incorporates 
quasi-bound quark-model states. Using the quark wave functions and the 
quark-meson interaction from the Cloudy Bag Model, we obtain 
a good overall agreement with the available experimental results
for the partial widths of the $N(1535)$ and the $N(1650)$ resonances
as well as for the pion, eta and kaon  electroproduction amplitudes.
Our model is consistent with the $N(1535)$ resonance being dominantly
a genuine three-quark state  rather than a quasi-bound state of mesons
and baryons.}


\maketitle

\section{\label{sec:intro}Introduction}

In the recent few years there has been a growing interest to
understand the nature of the resonances in the second resonance
region, in particular of the $N(1440)$ and the two closely lying
resonances in the S11 partial wave, the $N(1535)$ and $N(1650)$.
The new experiments on photo and electroproduction of mesons 
accompanied by the coupled-channel analyses by different groups  
\cite{Inna09,MAID2007,Anisovich10,Arndt06,Giessen05,Giessen08,%
Giessen09,KaonMAID}
along with analyses of elastic and inelastic scattering data 
performed mostly before 1995
\cite{Zagreb10,Speth03,Vrana2000,Manley92} have provided 
valuable information on the properties of these resonances.

From the theoretical side there have been substantial efforts 
to understand the peculiar nature of the lightest of the S11 
resonances, the $N(1535)$, due to its position just above the 
$\eta N$ threshold and the large branching ratio to the 
$\eta N$ channel.
In the quark model this resonance appears as a mixture of two 
\underline{70}-plet states with spin 1/2 and 3/2, which can 
explain the large $\eta N$ branching ratio of the $N(1535)$ 
and the almost complete absence of this decay in the case of 
its orthogonal partner, the $N(1650)$ \cite{Hey75}.
The mixing originates from the gluon and the meson interaction with 
quarks, and has been investigated in the constituent quark model 
\cite{Isgur77} and in the bag model \cite{deGrand76b,Myhrer84b}.
In an alternative picture, this resonance has been explained as 
a quasi-bound state of mesons and baryons with a strong admixture 
of the $K\Lambda$ and $K\Sigma$ (hidden) channels, in the framework 
of a potential model \cite{Siegel95} and the chiral perturbation 
theory \cite{Kaiser97,Arriola01,Oset02,Garcia04,Jido08,Mai11}.
The scattering amplitudes can be as well reproduced in 
coupled-channel dynamical models with genuine quark states 
for the two lowest S11 wave resonances 
\cite{JuliaDiaz06,JuliaDiaz07,Durand08} 
as in models assuming a quasi-bound state of mesons and baryons.
However, in the latter approaches, the calculation of meson 
photoproduction amplitudes \cite{Doering10} shows that an admixture 
of a genuine state corresponding to the $N(1650)$
is necessary in order to obtain the appropriate phases for 
the photoproduction amplitudes.
This shows that the nature of the $N(1535)$ is governed by a subtle
interplay of quark and meson degrees of freedom which requires
further investigations in coupled-channel dynamical models that 
involve quarks and mesons.

In our previous work we have developed a model that incorporates 
excited baryons represented as quasi-bound quark-model states into a 
coupled-channel formalism using the $K$-matrix approach \cite{EPJ2008}. 
In our method, the meson-baryon and the photon-baryon vertices are 
determined by the underlying quark model rather than fitted to the 
experimental data as is the case in phenomenological approaches.
We have investigated the P33 and P11 scattering as well as
electroproduction amplitudes dominated by the low-lying 
positive-parity resonances $\Delta(1232)$, $\Delta(1600)$ and 
$N(1440)$ \cite{EPJ2008,EPJ2009}.

The extension of the approach to low-lying negative-parity 
resonances requires the inclusion of new channels involving 
the $s$- and $d$-wave pions, the $\eta$ and the $\rho$ 
mesons, and the $K\Lambda$ channel.
The photoproduction of mesons has been studied in a chiral 
constituent quark model 
\cite{Saghai01,Saghai08e,Saghai08p,Saghai09,Saghai10}.
In this approach the widths of the resonances have been either
fitted or taken from the PDG while in our approach they are 
calculated from the quark-model wave functions.

In our previous calculation \cite{EPJ2008,EPJ2009} we have found 
a good agreement with the experiment for the scattering amplitudes 
as well as for the electroproduction amplitudes by using a simple 
chiral quark model, the Cloudy Bag Model (CBM) with the standard 
choice of parameters that had been widely adopted in studying the 
static properties of the nucleon and some low lying resonances.

The aim of this paper is to check whether the same model is able 
to describe a rather specific situation in the S11 partial wave 
with two closely lying resonances and a strong admixture of the 
inelastic channel very close to its threshold.
Furthermore, our aim is
to investigate whether the particular form of the quark-meson 
coupling that works well for the $p$-wave pions performs equally 
well for the $s$- and $d$-wave pions, and for the other pseudoscalar 
octet mesons.
Moreover,
we study the role of the meson cloud which  -- based on our 
experience in the P11 and P33 partial waves -- is expected to 
enhance rather 
substantially the strength of the meson-baryon 
and photon-baryon vertices  with respect to their bare values.

In the next section we briefly explain how our method can be
used to calculate the meson scattering and the electroproduction
amplitudes in a unified approach.
We generalize the approach of \cite{EPJ2009} to include the
production of mesons other than pions.

In sect.~3 we discuss the results for the scattering amplitudes
and concentrate on the $\eta N$ and $K\Lambda$ inelastic channels.
We investigate the sensitivity of the amplitudes in the resonance
region to the relative strength of the quark-pion and the 
quark-$\eta$ coupling.

In sect.~4 we present the results for the transverse helicity 
amplitude $A_{1/2}$ and the scalar helicity amplitude $S_{1/2}$ 
for the $N(1535)$ and $N(1650)$ resonances.
Finally, we present the results for the $\gamma N\to \pi N$
amplitude at the photon point and at $Q^2=1$~(GeV/$c$)$^2$,
and for photoproduction in the $\gamma N\to \eta N$
and  $\gamma N\to K\Lambda$ channels.

\section{\label{sec:basics}Basics of the $K$-matrix approach}

We consider a class of chiral quark models in which mesons
couple linearly to the quark core:
\begin{eqnarray}
 H' &=& 
  \int\d k \sum_{lmt}\left\{\omega_k\,a^\dagger_{lmt}(k)a_{lmt}(k)
\right. \nonumber\\ && \left.
    + \left[V_{lmt}(k) a_{lmt}(k) 
        + V_{lmt}^\dagger(k)\,a^\dagger_{lmt}(k)\right] \right\},
\label{Hpi}
\end{eqnarray}
where $a^\dagger_{lmt}(k)$ is the creation operator for a meson 
with angular momentum $l$, its third component $m$, and isospin 
$t$ (absent in the case of $s$-waves and isoscalar mesons).
Here $V_{lmt}(k)$ is a general form of the meson source 
involving the quark operators and is model dependent.
In appendix~\ref{vertices} we give a few examples for 
$V_{lmt}(k)$ in the Cloudy Bag Model.

We have shown  \cite{EPJ2008} that in such models the elements 
of the $K$~matrix in the basis with good total angular momentum 
$J$ and isospin $T$  take the form:
\begin{eqnarray}
 K_{M'B'\,MB}^{JT} &=&  -\pi\mathcal{N}_{M'B'}
   \langle\Psi^{MB}_{JT}||V_{M'}(k)||\widetilde{\Psi}_{B'}\rangle\,,
\nonumber\\
   \mathcal{N}_{MB} &=& \sqrt{\omega_{M} E_{B} \over k_{M} W}\,,
\label{defK}
\end{eqnarray}
where  $\omega_{M}$ and $k_{M}$ are the energy and momentum 
of the incoming or outgoing meson, $E_B$ is the baryon energy 
and $W$ is the invariant energy of the meson-baryon system.
The channels are also specified by the relative
angular momentum of the meson-baryon system and parity.
Here $|{\Psi}^{MB}\rangle$ is the principal-value state:
\begin{eqnarray}
|\Psi^{MB}_{JT}\rangle &=& \mathcal{N}_{MB}\left\{
    [a^\dagger(k_M)|\widetilde{\Psi}_B\rangle]^{JT} 
+
 \sum_{\mathcal{R}}c_{\mathcal{R}}^{MB}|\Phi_{\mathcal{R}}\rangle
\right. \nonumber\\ &+& \left.\kern-6pt 
 \sum_{M'B'}
   \int {\d k\>
       \chi^{M'B'MB}(k,k_M)\over\omega_k+E_{B'}(k)-W}\,
      [a^\dagger(k)|\widetilde{\Psi}_{B'}\rangle]^{JT}\right\}.
\kern12pt
\label{PsiH}
\end{eqnarray}
The first term represents the free meson ($\pi$, $\eta$, $\rho$, 
$K$, $\dots$) and the baryon ($N$, $\Delta$, $\Lambda, \ldots$) 
and defines the channel, the next term is the sum over 
{\em bare\/} three-quark states $\Phi_{\mathcal{R}}$ involving 
different excitations of the quark core, the third term 
introduces meson clouds around different isobars, $E(k)$ 
is the energy of the recoiled baryon.
In our approach we assume the commonly used picture in which 
the two-pion decay proceeds either through an unstable meson 
($\rho$-meson, $\sigma$-meson, \ldots) or through a baryon 
resonance ($\Delta(1232)$, $N(1440)$\, \ldots).
In such a case the state $\Psi^{MB}$ depends on the invariant 
mass of the subsystem (either $\pi\pi$ or $\pi N$) and the sum 
over $M'B'$ in (\ref{PsiH}) also implies the integration over 
the invariant mass.
The state $\widetilde{\Psi}_{B}$ is the asymptotic state of the
incoming or outgoing baryon; in the case when it corresponds to 
an unstable baryon, it depends on the invariant mass of the 
$\pi N$ subsystem, $M_B$, and is normalized as 
$\langle\widetilde{\Psi}_B(M_B')|
\widetilde{\Psi}_B(M_B)\rangle=\delta(M_B'-M_B)$,
where $M_B$ is the invariant mass of the $\pi N$ subsystem.
The meson amplitudes $\chi^{M'B'\,MB}(k,k_M)$ are proportional to 
the half off-shell matrix elements of the $K$-matrix
\begin{equation}
   K_{M'B'\,MB}(k,k_M)  = \pi\,\mathcal{N}_{M'B'}\mathcal{N}_{MB}\,
             \chi^{M'B'\,MB}(k,k_M) 
\label{chi2K}
\end{equation}
and obey an equation of the Lippmann-Schwinger type:
\begin{eqnarray}
&&   \chi^{M'B'\,MB}(k,k_M) 
   = -\sum_{\mathcal{R}}{c}^{MB}_{\mathcal{R}}\, {V}^{M'}_{B'\mathcal{R}}(k)
\nonumber\\ && 
       +\>\> \mathcal{K}^{M'B'\,MB}(k,k_M)
\nonumber\\ && 
+ \sum_{M''B''}\int\d k'\,
  {\mathcal{K}^{M'B'\,M''B''}(k,k')\chi^{M''B''\,MB}(k',k_M)
  \over \omega_k' + E_{B''}(k')-W}\,,
\nonumber\\
\label{eq4chi}
\end{eqnarray}
where
\begin{equation}
  {\cal K}^{M'B'\,MB}(k,k')
=  
  \sum_{B''} f_{BB'}^{B''}\,
  {\widetilde{\mathcal{V}}_{B''B'}^{M'}(k') \, 
   \widetilde{\mathcal{V}}_{B''B}^{M}(k)
   \over \omega_k+\omega_k'+E_{B''}(\bar{k})-W}\,,
\label{kernel}
\end{equation}
\begin{eqnarray}
  f_{AB}^C  &=& \sqrt{(2J_A+1)(2J_B+1)(2T_A+1)(2T_B+1)}\,
\nonumber \\
&\times&   W(l_BJ_AJ_Bl_A;J_C,J)W(i_BT_AT_Bi_A;T_C,T)\,,
\label{fabc}
\end{eqnarray}
and $l_A$, $i_A$, $J_A$, and $T_A$ are the meson and 
the baryon angular momenta and isospins, respectively.
The coefficients ${c}^{MB}_{\mathcal{R}}$ obey the equation
\begin{eqnarray}
 (W-M_{\mathcal{R}}^{(0)}) {c}^{MB}_{\mathcal{R}}
   &=& V^M_{B\mathcal{R}}(k_M) 
\nonumber\\
&+& \sum_{M'B'}\int\d k\,
    {\chi^{M'B'\,MB}(k,k_M) V^{M'}_{B'\mathcal{R}}(k)\over 
     \omega_k + E_{B'}(k)-W}\,.
\nonumber\\
\label{eq4c}
\end{eqnarray}
Here ${V}^{M}_{B\mathcal{R}}(k)$ are the matrix elements of the 
quark-meson interaction between the baryon state $B$ and the
bare three-quark state $\Phi_{\mathcal{R}}$, and $M_{\mathcal{R}}^{(0)}$ 
is the energy of the bare state.
Solving the coupled system of equations (\ref{eq4chi}) and 
(\ref{eq4c}) using a separable approximation for the kernels 
(\ref{kernel}) (see  \cite{EPJ2008}), the resulting amplitudes 
take the form
\begin{equation}
   \chi^{M'B'MB}(k,k_M) 
     = -\sum_{\mathcal{R}}\widetilde{c}^{MB}_{\mathcal{R}}\,
    \widetilde{\cal V}^{M'}_{B'\mathcal{R}}(k)
       + \mathcal{D}^{M'B'MB}(k,k_M)\,,
\label{sol4chi}
\end{equation}
where the first term represents the contribution of
various resonances while $\mathcal{D}^{M'B'\,MB}(k)$ 
originates in the non-resonant background processes. 
The physical resonant state $\mathcal{R}$ is a superposition 
of the dressed states built around the bare three-quark states 
$\Phi_{\mathcal{R}'}$.
Here
\begin{equation}
    \widetilde{c}_{\mathcal{R}}^{MB} 
          = {\widetilde{\cal V}^M_{B\mathcal{R}}
             \over Z_{\mathcal{R}}(W) (W-M_{\mathcal{R}})}\,,
\label{calVmix}
\end{equation}
$Z_{\mathcal{R}}$ is the wave-function normalization, 
$\widetilde{\mathcal{V}}^M_{B\mathcal{R}}$ is the dressed matrix 
element of the quark-meson interaction between the resonant state 
and the baryon state in the channel $MB$, and obeys the equation 
\begin{equation}
\widetilde{\cal V}^M_{B\mathcal{R}}
   = {V}^{M}_{B\mathcal{R}}(k)
+ \sum_{M'B'}\int\d k'\,
  {\mathcal{K}^{MB\,M'B'}(k,k')\,\widetilde{\cal V}^{M'}_{B'\mathcal{R}}(k')
  \over \omega_k' + E_{B'}(k')-W}\,.
\label{eq4calV}
\end{equation}
The non-resonant background amplitude obeys
\begin{eqnarray}
&&   \mathcal{D}^{M'B'\,MB}(k,k_M)
       = \mathcal{K}^{M'B'\,MB}(k,k_M)
\nonumber\\ && 
+ \sum_{M''B''}\int\d k'\,
  {\mathcal{K}^{M'B'\,M''B''}(k,k')\mathcal{D}^{M''B''\,MB}(k',k_M)
  \over \omega_k' + E_{B''}(k')-W}\,.
\nonumber\\
\label{eq4D}
\end{eqnarray}
Let us mention that the resulting set of integral equations is equivalent 
to the set of equations for the dressed vertices and the non-pole 
part of the $T$ matrix (see e.g. \cite{Doering09}, eq. (4)).
In \cite{EPJ2008} we have described the numerical procedure
using a separable approximations for the kernels (\ref{kernel})
to solve the system.

The $T$ matrix is finally obtained by solving the Heitler's 
equation
\begin{equation}
T = K +\i\, TK\,.
\label{Heitler}
\end{equation}

The method can be extended in a straightforward manner 
to the calculation of electroproduction amplitudes by
including the $\gamma N$ channel.
As we have shown in \cite{EPJ2009} the electroproduction
amplitudes for the pion can be split into the resonant part
and the background part.
This splitting can be generalized to electroproduction
of other mesons.
The resonant part reads
\begin{equation}
{\mathcal{M}_{\gamma N MB}^\mathrm{(res)}}  = 
\sqrt{\omega_\gamma E_N^\gamma \over \omega_M E_B }\,
{\zeta\over\pi\widetilde{\cal V}_{BN^*}}\,
   \langle\Psi_{N^*}^\mathrm{(res)}(W)|{\tilde{V}_\gamma}
                 |\Psi_N\rangle\, {T_{MB\,MB}} \>,
\label{Mres}
\end{equation}
where ${\tilde{V}_\gamma(\mu,\vec{k}_\gamma)}$ is the interaction 
of the photon with the electromagnetic current which contains
quark and pion contributions,
$$
\vec{j}_{EM}(\vec{r})  =
  \bar{\psi}\vec{\gamma}({\textstyle{1\over6}} + \half\tau_0)\psi
  + \i \sum_t t \pi_t(\vec{r})\vec{\nabla}\pi_{-t}(\vec{r})\,,
$$
$T_{MB\,MB}$ is the meson-baryon scattering amplitude, and $\zeta$ 
is the spin-isospin factor depending on the considered multipole 
and the spin and isospin of the outgoing hadrons.
The state $\Psi_{N^*}^{\mathrm{(res)}}$ is obtained from (\ref{PsiH}) 
by keeping only the linear combination of ${\cal R}$ representing 
$N^*$ in the sum over ${\cal R}$ and the corresponding term in the
sum in (\ref{sol4chi}).  
The expression 
$\langle\Psi_{N^*}^\mathrm{(res)}(W)|\tilde{V}_\gamma|\Psi_N\rangle 
= A_{\gamma N\to N^*}$ 
is the electroexcitation amplitude of the chosen resonance $N^*$.

The background part of the electroproduction amplitude satisfies
\begin{equation}
\mathcal{M}_{MB}^\mathrm{(bkg)} = 
       \mathcal{M}_{MB}^{K\,\mathrm{(bkg)}}
      +\i\sum_{M'B'}T_{MB\,M'B'}\mathcal{M}_{M'B'}^{K\,\mathrm{(bkg)}}. 
\label{Mnon}
\end{equation}
Here $\mathcal{M}_{M'B'}^{K\,\mathrm{(bkg)}}$ is the expectation value
of the electromagnetic interaction between the nucleon and the 
channel state $\Psi_{M'B'}^{JT}$ (\ref{PsiH}) without the terms 
pertinent to the resonant state $N^*$.
It contains the contribution from the other resonances, from the
scattering background (the ${\cal D}$ term in (\ref{sol4chi})),
and from the first term in (\ref{PsiH}).
The latter term, representing the unperturbed meson, is
responsible for the terms corresponding to the $t$- and 
$u$-channel processes.


\section{\label{sec:scattering}The S11 scattering amplitudes}

We have used the Cloudy 
Bag Model to describe the quark wave-functions and quark-meson 
coupling; the details are given in appendix~\ref{vertices}.
The parameters of the model are the bag radius and the strength 
of the quark-pion coupling $f_\pi$ (corresponding to the pion 
decay constant) set to $f_\pi=76$~MeV  which reproduces 
the experimental value of the $\pi NN$ coupling constant.
For the bag radius we use $R_\mathrm{bag}=0.83$~fm yielding best 
results for the ground state properties and also for the 
scattering and electroproduction amplitudes in the P11 and 
P33 partial waves.
Further free parameters of the present calculations are 
the masses of the bare three-quark states and the mixing 
angle $\vartheta_S$ discussed below.

To calculate the scattering and electroproduction amplitudes 
in the S11 partial wave we have included the $\pi N$, 
$\pi\Delta(1232)$, $\pi N(1440)$, $\rho N$, and $K\Lambda(1116)$ 
channels, and the $N(1535)$ and $N(1650)$ resonances.
The quark-model wave-functions for the negative-parity states in 
the $j$-$j$ coupling scheme have been taken from \cite{Myhrer84b}:
\begin{eqnarray}
\Phi_{\cal R} 
  &=& c^{\cal R}_{A} \vert (1s)^2(1p_{3/2})^1 \rangle
    + c^{\cal R}_{P} \vert (1s)^2(1p_{1/2})^1 \rangle_1
\nonumber\\ &&
    + c^{\cal R}_{P'} \vert (1s)^2(1p_{1/2})^1 \rangle_2\,,
\label{calR}
\end{eqnarray}
where the mixing coefficients $c^{\cal R}_{A}$, $c^{\cal R}_{P}$, 
and $c^{\cal R}_{P'}$ are expressed in terms of the mixing angle 
$\vartheta_s$ between the spin-$1/2$ and spin-$3/2$ 
three-quark configurations. 
They are given by
\begin{eqnarray*}
c^1_{A} &=& -{1\over3}(2\cos\vartheta_s +\sin\vartheta_s), 
\nonumber\\ 
c^1_{P} &=&  {\sqrt2\over6}(\cos\vartheta_s -4\sin\vartheta_s) , 
\quad 
c^1_{P'} =  {\sqrt2\over2}\,\cos\vartheta_s\,,
\nonumber\\ 
c^2_{A} &=& {1\over3}(\cos\vartheta_s - 2\sin\vartheta_s), 
\nonumber\\ 
c^2_{P} &=&  {\sqrt2\over6}(\sin\vartheta_s + 4\cos\vartheta_s), 
\quad 
c^2_{P'} =  {\sqrt2\over2}\,\sin\vartheta_s \,.
\end{eqnarray*}
In this coupling scheme the spurious translational state belonging
to the SU(6) spin-flavour $\underline{56}$-plet is projected out.

The mixing is a consequence of the gluon and the meson 
interactions; since the quark-gluon interaction is not included 
in the model, the mixing angle due to the gluons is taken as 
a free parameter independent of $W$.
In the energy region of the $N(1535)$ and $N(1650)$ resonances 
we obtain the best results using $\vartheta_s=-34^\circ$  for the 
bare states; after taking into account the mixing due to the 
meson loops, the value of the mixing angle reaches  $-33^\circ$ 
in agreement with the phenomenological analysis \cite{Hey75}.
The results, however, do not depend strongly on the mixing angle,
and deviations in $\vartheta_s$ of the order $\approx 5^\circ$
lead to similar results.
The poles of the $K$ matrix are at 1535~MeV and 1690~MeV, and
almost exactly coincide with zero crossings of the real parts
of the $\eta N$ and $\pi N$ amplitudes, respectively.
The corresponding masses of the bare three-quark states are
1720~MeV and 1815~MeV, which are close to the bare values 
used in \cite{JuliaDiaz07}.

\begin{figure}[h!]
\begin{center}
\includegraphics[height=40mm]{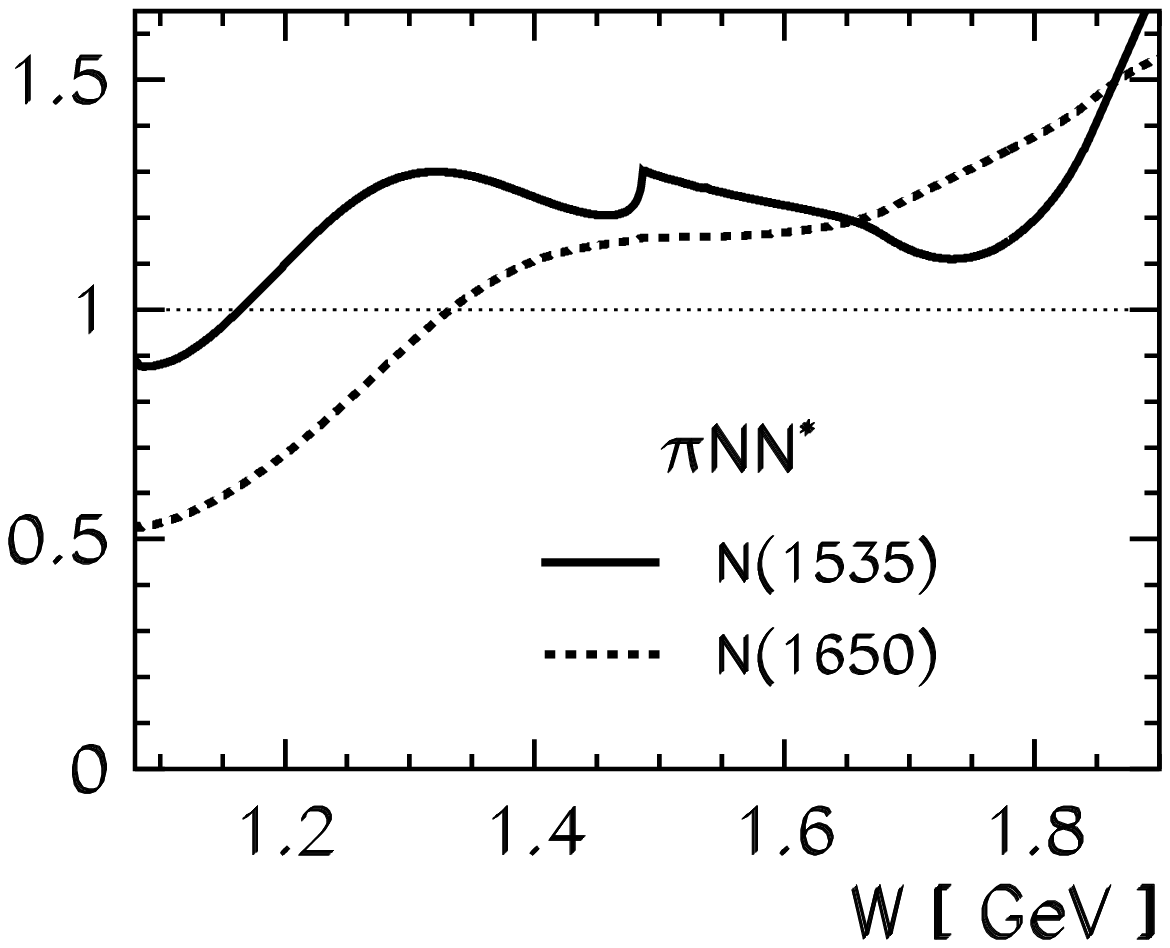}
\includegraphics[height=40mm]{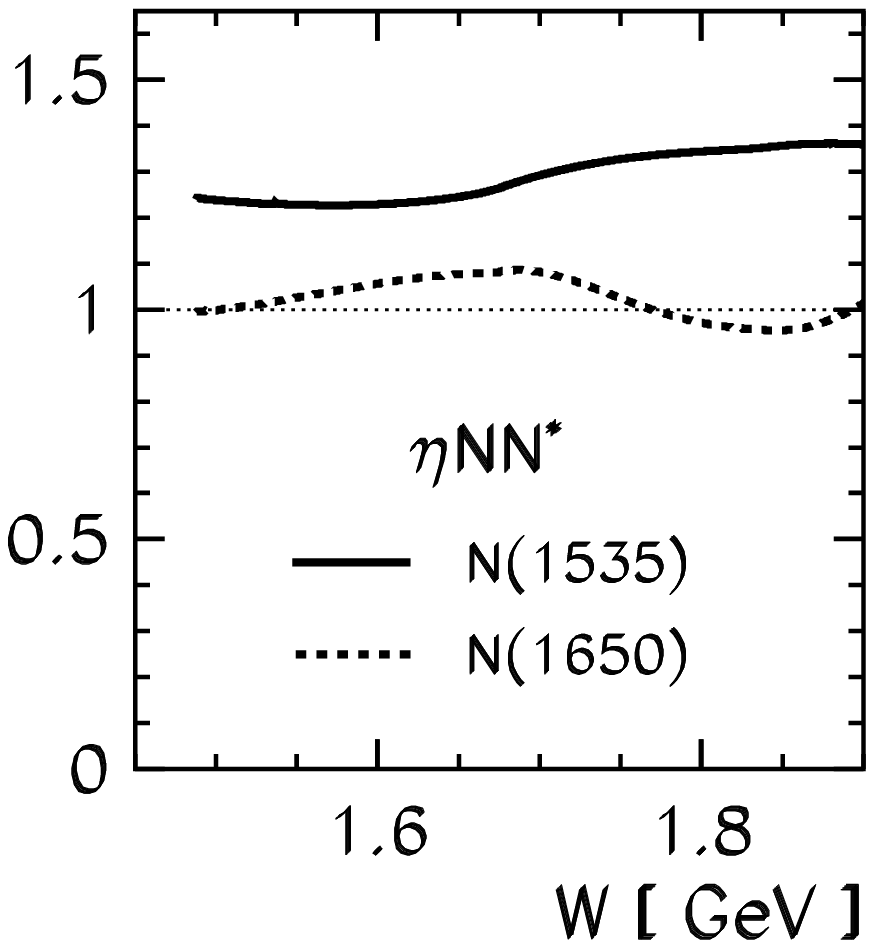}
\end{center}
\caption{Left panel: the ratio of the dressed and the bare 
$\pi NN^*$ vertex for the $N(1535)$ and the $N(1650)$; 
right panel: same for the $\eta NN^*$ vertex.}
\label{fig:dressVert}
\end{figure}

To generate the kernel (\ref{kernel}) of the equations for
the dressed vertex (\ref{eq4calV}) and the non-resonant background
(\ref{eq4D}) we have taken into account the $s$- and $p$-wave pions,
the $s$-wave $\eta$ mesons, and the following intermediates states:
$N$, $N(1440)P_{11}$, $\Delta(1232)P_{33}$, $\Delta(1600)P_{33}$,
$N(1535)S_{11}$,\break
$N(1650)S_{11}$, $N(1520)D_{13}$,  $N(1700)D_{13}$, 
$\Delta(1630)D_{31}$,\break  
$\Delta(1700)D_{33}$.
All baryon-meson couplings have been calculated in the Cloudy
Bag Model while the standard PDG values \cite{PDG} have been 
used for the masses of the intermediate states.
The contributions from the $d$-wave pions and $J={5\over2}$ 
resonances as well as from the $s$-wave kaons turn out to be small.  
Figure~\ref{fig:dressVert} shows the dressed vertices
$\widetilde{\cal V}^\pi_{NN^*}(k_\pi)$ and 
$\widetilde{\cal V}^\eta_{NN^*}(k_\eta)$ for $N^*$ corresponding to
the $N(1535)$ and $N(1650)$ divided by the corresponding bare values.
The enhancement in the resonance region is smaller than
in the case of the P11 and the P33 partial wave \cite{EPJ2008}.

{\renewcommand{\arraystretch}{1.3}
\setlength{\tabcolsep}{5pt}
\begin{table}[h]
\begin{center}
\begin{tabular}{|l|c|c|c|c|c|c|c|}
\hline

\hline
Res.       & $\Gamma_\mathrm{tot}$ 
          &\multicolumn{6}{|c|}{$\Gamma_i/\Gamma_\mathrm{tot}$} \\ 
\hline
 &  [MeV] & $\pi N$ & $\eta N$ & $\pi\Delta$ 
   & $K\Lambda$ & $\rho_1N$ & $\pi R$ \\
\hline
$N(1535)$  &  95     & 0.44    & 0.54 & 0.01 & -    & 0.02 & 0    \\
$N(1535)$' & 128     & 0.35    & 0.62 & 0.01 & -    & 0.01 & 0    \\
Exp.       & 125     & 0.35    & 0.53 & 0.01 & -    & 0.02 & 0    \\[-4pt]
           &  -- 175 & -- 0.55 &      &      &      &      &      \\
\hline
$N(1650)$  & 144     & 0.60    & 0.01 & 0.22 & 0.09 & 0.04 & 0.04 \\
$N(1650)$' & 124     & 0.71    & 0.01 & 0.09 & 0.10 & 0.05 & 0.04 \\
Exp.       & 150     & 0.60    & 0.02 & 0.02 & 0.03 & 0.01 & 0.03 \\[-4pt]
           &  -- 180 & -- 0.95 &      &      &      &      &      \\
\hline

\hline
\end{tabular}
\end{center}
\caption{The total and the partial widths for the $N(1535)$ 
and the $N(1650)$ resonance at the $K$-matrix pole (1535~MeV 
and 1690~MeV, respectively). 
Here $N(1535)$' stands for the results using the value 
$f_\eta=f_\pi$ (instead of $f_\eta=1.2\,f_\pi$), and $N(1650)$' for 
the results using the  value of the  $d$-wave  $\pi\Delta$ 
coupling reduced by a factor of 0.50 compared to the quark-model 
prediction; $\pi R$ means the $\pi N(1440)$ channel.
The experimental values are from \cite{PDG}.}
\label{bg:tableS11}
\end{table}}

\begin{figure}[h!]
\begin{center}
\includegraphics[width=80mm]{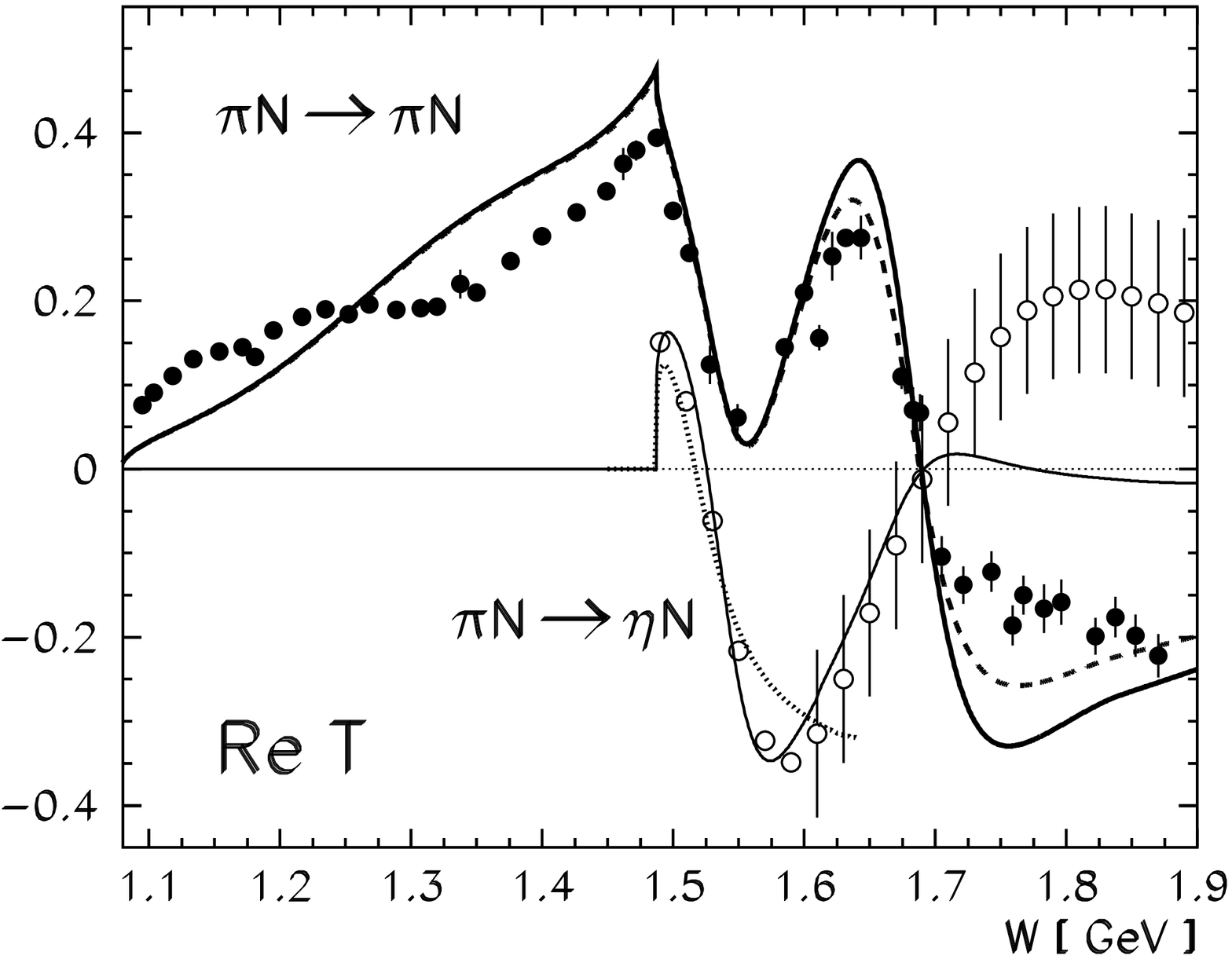}
\includegraphics[width=80mm]{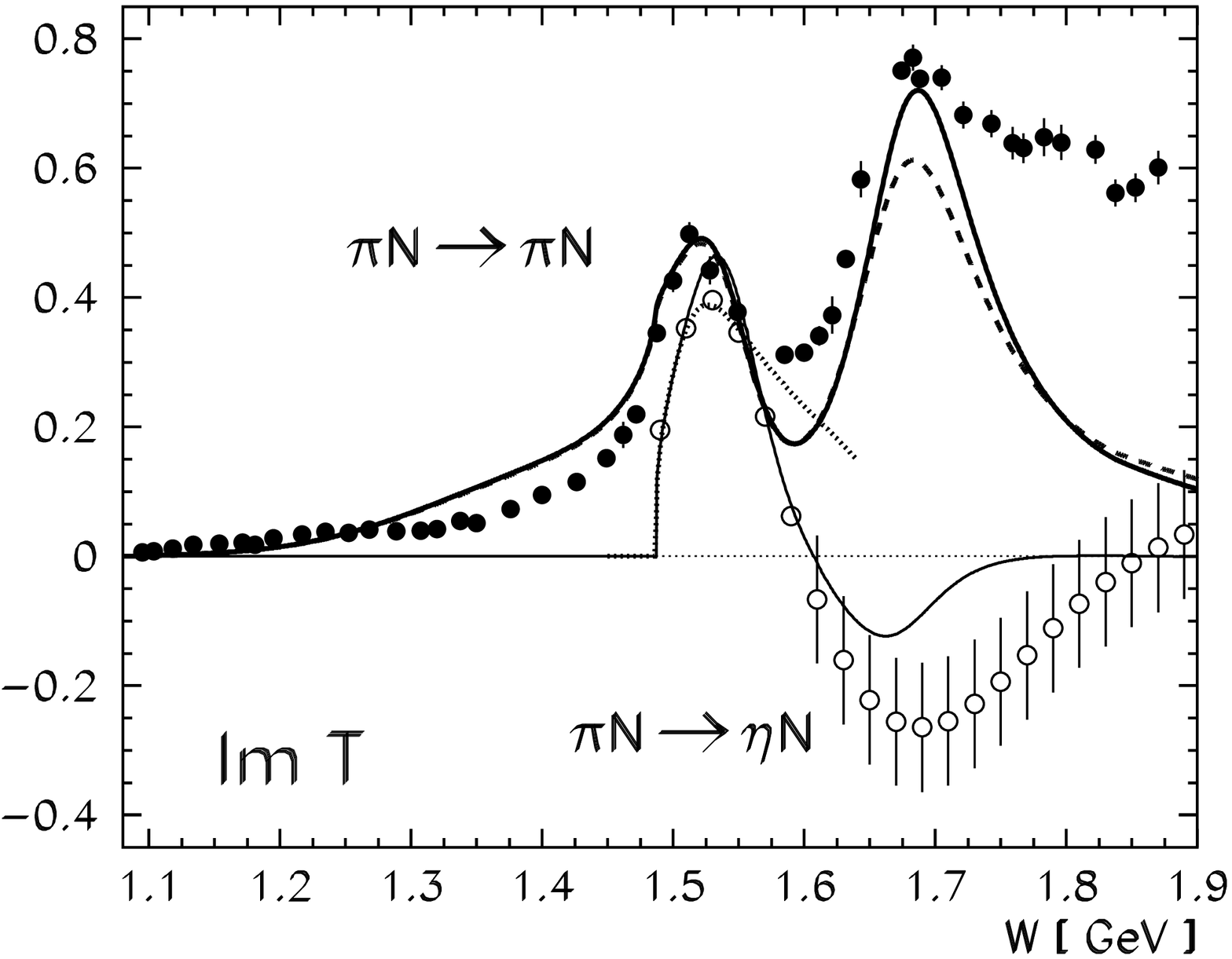}
\end{center}
\caption{The real and the imaginary part of the scattering
$T$ matrix for the S11 partial wave.
The thick solid and dashed lines correspond to the elastic 
channel and the thin solid and the thin dotted lines to 
the $\pi N \rightarrow\eta N$ channel.
The solid lines are obtained by using the reduced value of the 
$d$-wave $\pi\Delta$ coupling while the dashed lines correspond to 
the unmodified quark-model values for the baryon-meson couplings.
The points for the elastic channel (full circles) are from the 
SAID $\pi N\to\pi N$ partial-wave analysis \protect\cite{Arndt06}. 
Those for the inelastic one (open circles) 
are from \cite{Capstick04}; the dotted lines are the latest (WI08)
$\pi N \rightarrow\eta N$ solution \cite{Strakovsky}.}
\label{fig:S11pieta}
\end{figure}

\begin{figure}[h!]
\begin{center}
\includegraphics[width=\hsize]{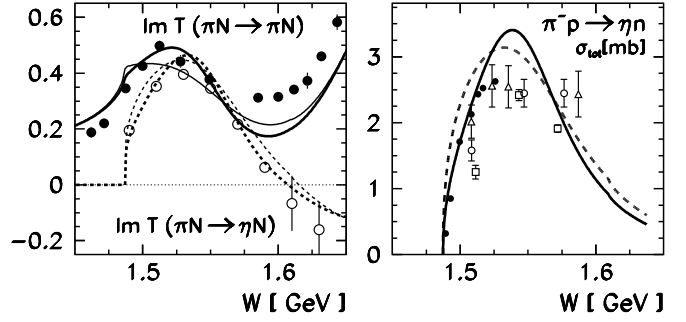}
\end{center}
\caption{Left panel: the imaginary part of the $T$ matrix for 
elastic scattering (solid lines) and for the $\eta N$ channel 
(dashed lines) in the region of the $N(1535)$.
The thick lines correspond to $f_\eta=1.2\,f_\pi$, 
the thin lines to $f_\eta=f_\pi$.
Right panel: the (dominant) S11 contribution to the total 
cross-section for the $\pi^-\,p\rightarrow \eta\, n$ reaction
using $f_\eta=1.2\,f_\pi$ (solid line) and $f_\eta=f_\pi$ 
(dashed line).
Data points in the left panel: see fig.~\ref{fig:S11pieta}.  
Experimental data in the right panel are from
\cite{Deinet69} (open triangles), \cite{Richards70} (open circles),
\cite{Brown79} (open squares), and \cite{Prakhov05} (filled circles).}
\label{fig:S11pietazoom}
\end{figure}

\begin{figure}[h!]
\begin{center}
\includegraphics[width=\hsize]{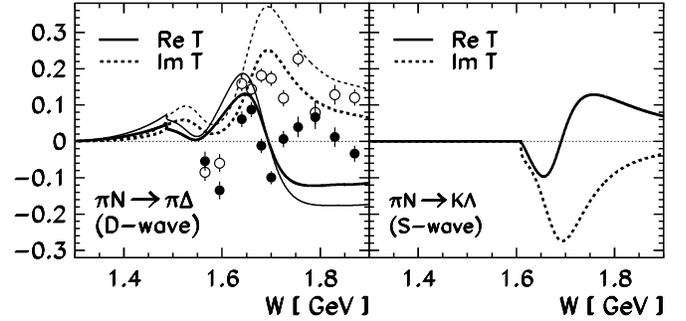}
\end{center}
\caption{The real and the imaginary part of the $T$ matrix 
for the $\pi N \rightarrow \pi\Delta$ (left panel) and for  
the $\pi N \rightarrow K\Lambda$ channel (right panel)  
in the S11 partial wave.
The thick lines are obtained by using the reduced value of the 
$d$-wave $\pi\Delta$ coupling while the thin lines correspond to 
the unmodified quark-model values for the baryon-meson couplings.
The points are from the partial-wave analysis of \cite{Manley92}.}
\label{fig:S11ch}
\end{figure}

The results for pion-induced meson production are displayed in 
table~\ref{bg:tableS11} and 
figs.~\ref{fig:S11pieta}--\ref{fig:S11ch}.
In the vicinity of the lower ($N(1535)$) resonance, just above 
the $\eta$ threshold, the elastic and inelastic amplitudes 
are strongly influenced by the $s$-wave $\eta N$ channel. 
In the energy region of the upper resonance ($N(1650)$), 
additional channels open or become more important.
We have considered the following additional channels: the 
$\pi\Delta$ channel with $l=2$, the $K\Lambda$ channel with $l=0$, 
two channels involving the $\rho$ meson with $l=0$ ($\rho_1N$) 
and $l=2$ ($\rho_3N$), and the $\pi N(1440)$ channel with $l=0$.
Using the quark-model values for the quark-meson coupling as 
introduced in appendix~\ref{vertices} we obtain a good agreement 
between the model prediction and the experimental analysis for 
the lower resonance; for the upper resonance the agreement is worse.
Though the extraction of the experimental points is less reliable 
and differs considerably between different authors, the results 
clearly indicate that the strength of the $\pi\Delta$ ($d$-wave) 
vertex is overestimated in our model (table~\ref{bg:tableS11}).
Multiplying the strength of the  $\pi\Delta$ vertex by 0.5 
yields a better agreement with the data (fig.~\ref{fig:S11ch})  
and improves the agreement with the imaginary part of the 
elastic $T$ matrix (fig.~\ref{fig:S11pieta}).

The behaviour of the imaginary part of the $\pi N$ amplitude is 
very sensitive to the value of $f_\eta$; for our standard choice 
$f_\eta=1.2 f_\pi$ we get a typical resonant peak at around 1510 MeV 
while for $f_\eta=f_\pi$ (corresponding to a stronger quark-$\eta$
coupling) this peak disappears and we end up with a cusp at the 
$\eta N$ threshold (fig.~\ref{fig:S11pietazoom}).
The effect of reducing $f_\eta$ also considerably influences 
the total width of the resonances and the $\pi N$ branching ratio
(table~\ref{bg:tableS11}).
This effect is not so pronounced in the case of the $\eta N$ 
amplitudes.
The comparison of the calculated total cross section for the
reaction $\pi^- p \rightarrow \eta\, n$ which is dominated
by the $\eta N$ channel up to $W\sim 1650$~MeV \cite{Mosel02}
with the recent precise measurement \cite{Prakhov05} seems to
favour the higher value of $f_\eta$ (fig.~\ref{fig:S11pietazoom}, 
right panel).

The bag radius determines the overall strength of the quark-meson
interaction and an effective momentum cut-off which for the
$s$-wave mesons and $R_\mathrm{bag}=0.83$~fm corresponds to
$\Lambda = 510$~MeV/$c$.
Taking a larger radius, $R_\mathrm{bag}=0.90$~fm, the widths of the 
resonances are reduced by $\sim 35$~\%, but the behaviour of the 
elastic amplitude improves in the region below the $\eta N$ threshold.
Decreasing the bag radius the widths increase and at the same time 
the mass of the bare $N(1535)$ increases while that of the $N(1650)$ 
decreases.
Below a certain critical $R_\mathrm{bag}$ no solution exists for 
the bare masses if the positions of the $K$-matrix poles remain
fixed; this might be an indication that the picture with two 
genuine three-quark states breaks down and quark-meson couplings 
become sufficiently strong to support a quasi-bound state of 
mesons and baryons.


\section{\label{eproduction}The meson electroproduction 
amplitudes}

In the S11 partial wave, electroproduction of mesons 
is dominated by the $E_{0+}$ and $S_{0+}$ amplitudes.
The corresponding $E1$ and $C1$ multipole operators have
the familiar forms
\begin{eqnarray}
V^{E1}_\gamma &=& 
  {\i\e\,\sqrt{3\pi}\,\mu\over\sqrt{2\omega_\gamma}\,k_\gamma}
   \int\d\vec{r}\,\biggl[\vec{j}_{EM}(\vec{r})\cdot
       \vec{\nabla}\left(Y_{1\mu}(\hat{r})
   {\partial\over\partial r}rj_1(k_\gamma r)\right)\biggr. 
\nonumber\\ && \biggl.
+ k_\gamma ^2\,\vec{r}\cdot
     \vec{j}_{EM}(\vec{r})\; j_1(k_\gamma r)\, Y_{1\mu}(\hat{r})\biggr]
\end{eqnarray}
and
\begin{equation}
 V^{C1}_\gamma = {\e\;\sqrt{12\pi}\over\sqrt{2\omega_\gamma}}\;
      \int\d\vec{r}\,j_1(k_\gamma r)Y_{10}(\hat{\vec{r}})
      \rho_{EM}(\vec{r})\,.
\end{equation}
The relations between the helicity and the pion electroproduction 
amplitudes are the same as for the $M1$ multipole and are given
in \cite{EPJ2009}.
For $\zeta$ in (\ref{Mres}) we have $\zeta_E = \sqrt{1/12}$ for the 
transverse and  $\zeta_S = \sqrt{1/6}$ for the scalar amplitude, 
and $\zeta_E = \sqrt{1/2}$ and $\zeta_S = 1$ for the $\eta N$ and 
$K\Lambda$ amplitudes.

\begin{figure}[h!]
\begin{center}
\includegraphics[width=\hsize]{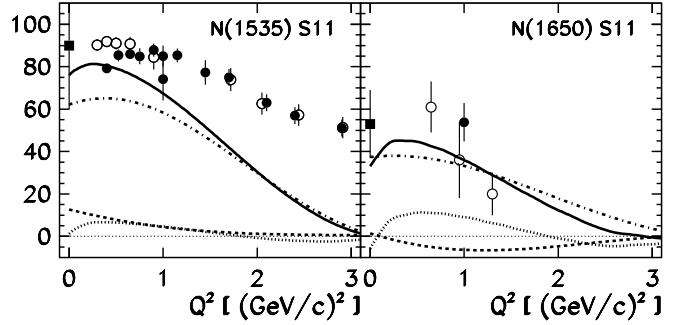}
\end{center}
\caption{The proton transverse helicity amplitudes (solid lines) 
for the $N(1535)$ (left panel) and for the $N(1650)$ (right panel)
evaluated at the pole of the $K$ matrix, in units of 
$10^{-3}~{\rm GeV}^{-1/2}$.
The dashed-dotted lines correspond to the quark contribution,
the dotted lines to the pion contribution and the dashed lines
to vertex corrections.
Experimental data are from the PDG \cite{PDG} (filled squares 
at $Q^2=0$), MAID \cite{MAID2007,MAID2009} (filled circles), and 
analyses of CLAS data \cite{Inna09,Mokeev10} (open circles).}
\label{fig:Thelicity}
\end{figure}

\begin{figure}[h!]
\begin{center}
\includegraphics[width=\hsize]{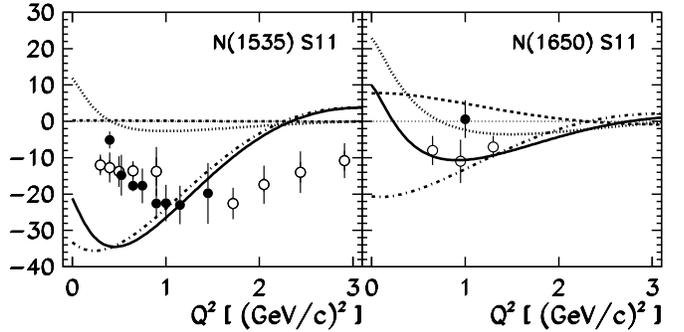}
\end{center}
\caption{The proton scalar helicity amplitudes for the
$N(1535)$ and  $N(1650)$ resonances.
Notation as in fig.~\ref{fig:Thelicity}.}
\label{fig:Shelicity}
\end{figure}

The helicity amplitudes are displayed in 
fig.~\ref{fig:Thelicity} and fig.~\ref{fig:Shelicity}.
The values for the transverse amplitudes are well 
reproduced at the photon point and in the low-$Q^2$ region.
For $Q^2>1.5$~(GeV/$c$)$^2$ the calculated amplitudes become
considerably smaller compared to the data indicating that our 
simple quark model is not able to reproduce the short range 
behaviour of the quark wave-function in the nucleon.
As expected, the pion cloud makes a sizable contribution
in the region of low $Q^2$.
In the ${N}(1535)$ case, the pion contribution and the 
vertex corrections improve the agreement with the data.
For the scalar amplitude this contribution is however not 
sufficient to partially cancel the large quark part. 
In the transverse amplitude for the ${N}(1650)$, the strength 
of the pion cloud contributions is about half of the quark 
part, while in the scalar case, the pion and quark parts 
largely cancel at low $Q^2$ and bring the sum into a better 
agreement with the data compared to the $N(1535)$ case.

The corrections for center-of-mass motion have been studied 
in the Cloudy Bag Model in the case of helicity amplitudes for 
the $\Delta(1232)$ production \cite{Thomas97}.
They reduce the values of the amplitudes at smaller $Q^2$ and
enhance them at larger $Q^2$.
However, the effect is small and we have not considered 
such corrections in the present calculation.

\begin{figure}[h!]
\begin{center}
\includegraphics[width=\hsize]{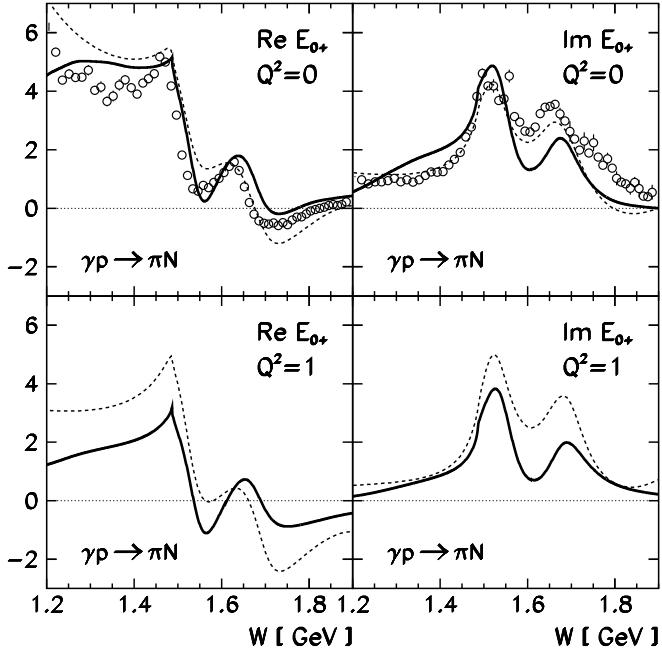}
\end{center}
\caption{The transverse  amplitude $E_{0+}$ for  
$\gamma p\rightarrow\pi N$ at $Q^2=0$ (upper panel) and 
$Q^2=1$~(GeV/$c$)$^2$ (lower panel) in units of 
$10^{-3} m_\pi^{-1}$.
Dashed lines correspond to the MAID analysis \cite{MAID2007},
the experimental data are from \cite{SAIDpi}.}
\label{fig:E0pipro}
\end{figure}

\begin{figure}[h!]
\begin{center}
\includegraphics[width=\hsize]{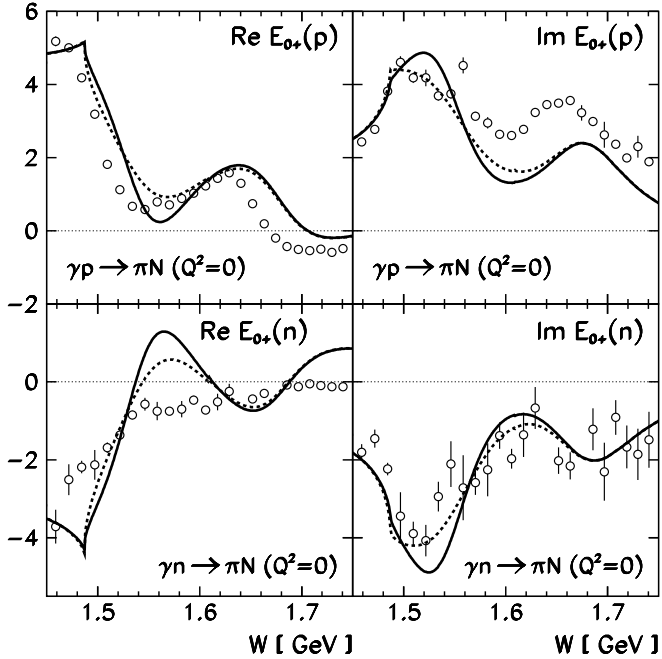}
\end{center}
\caption{The transverse pion photoproduction amplitudes $E_{0+}$ 
on the proton (upper panels, zoom-in of fig.~\ref{fig:E0pipro}) 
and on the neutron (lower panels) in units $10^{-3} m_\pi^{-1}$.
The solid lines correspond to $f_\eta=1.2\,f_\pi$, 
the dashed lines to $f_\eta=f_\pi$.
The experimental data are from \cite{SAIDpi}.}
\label{fig:E0pipn}
\end{figure}

\begin{figure}[h!]
\begin{center}
\includegraphics[width=\hsize]{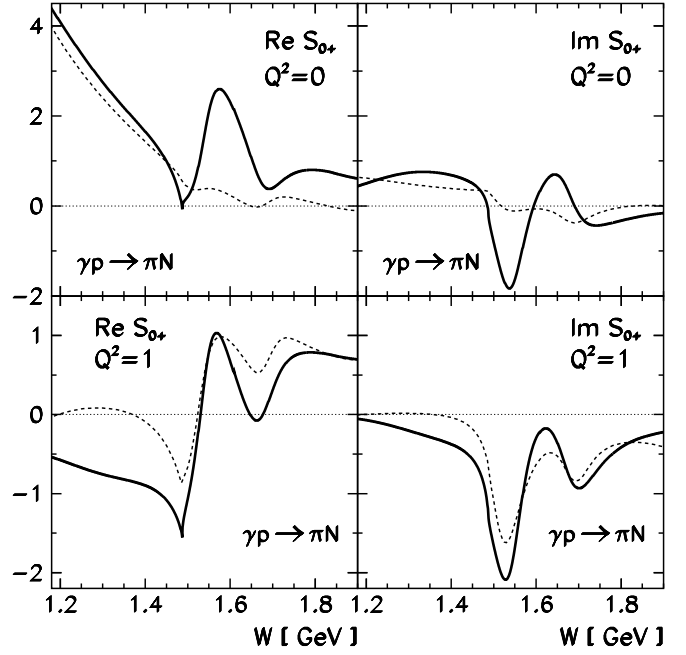}
\end{center}
\caption{The scalar pion production amplitude $S_{0+}$.
Notation as in fig.~\ref{fig:E0pipro}.}
\label{fig:S0pipro}
\end{figure}

The $E_{0+}$ electroproduction amplitude for 
$\gamma p\rightarrow\pi N$ is displayed in fig.~\ref{fig:E0pipro} 
and fig.~\ref{fig:E0pipn} for $Q^2=0$ and $Q^2=1$~(GeV/$c$)$^2$,
and for $\gamma n\rightarrow\pi N$ at $Q^2=0$ in 
fig.~\ref{fig:E0pipn}.
Since the model reproduces well the transverse helicity amplitudes 
in the resonance region, the electroproduction amplitudes also 
agree well with the phenomenologically determined amplitudes.
The good agreement extends also into the region below the resonance 
where the dominant contribution arises from the pion $t$-channel.
Our model, however, fails to reproduce the threshold behaviour of 
the $E_{0+}$ amplitudes since it assumes the pseudoscalar pion-quark 
coupling and hence does not contain the Kroll-Ruderman term.
(This term can be {\em ad hoc\/} included in the model but since 
we are primarily interested in the resonance region we have not 
considered such a possibility.)
The $S_{0+}$ scalar amplitude for $\gamma p\rightarrow\pi N$
is shown in  fig.~\ref{fig:S0pipro}.
Since our model overestimates the size of the scalar helicity 
amplitude at the photon point it also disagrees with the 
phenomenologically determined amplitudes in the resonance 
region for $Q^2=0$ but yields a consistent prediction at 
$Q^2\sim1$~(GeV/$c$)$^2$.

\begin{figure}[h!]
\begin{center}
\includegraphics[width=\hsize]{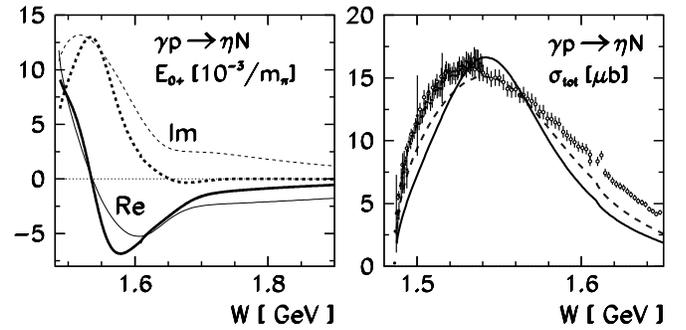}
\end{center}
\caption{Eta photoproduction on the proton.
Left panel: real and imaginary parts of the $E_{0+}$ multipole; 
our result (thick lines); EtaMAID \cite{EtaMAID} (thin lines).
Right panel: total cross-section; our result ($f_\eta=1.2\,f_\pi$: 
solid line, $f_\eta=f_\pi$: dashed line).  
The experimental data are from \cite{EtaSAID} and \cite{McNicoll}.}
\label{fig:E0etapro}
\end{figure}

The results for $\eta$ photoproduction are compared to 
the phenomenological prediction for the $E_{0+}$ amplitude 
and the total cross section in fig.~\ref{fig:E0etapro}.
The results almost entirely depend on the properties of the 
lower $N(1535)$ resonance and the threshold behaviour of the 
$\eta N$ amplitude but very little on the background processes.
Our standard choice $f_\eta=1.2\,f_\pi$ yields somewhat too 
narrow amplitudes; in contrast to the $\pi N\rightarrow \eta N$ 
process, the choice $f_\eta=f_\pi$ leads to a better agreement.
The good overall agreement with the data for $\eta$ production 
supports our conjecture about the dominance of the 
genuine three-quark configuration in the $N(1535)$ state.

\begin{figure}[h!]
\begin{center}
\includegraphics[width=\hsize]{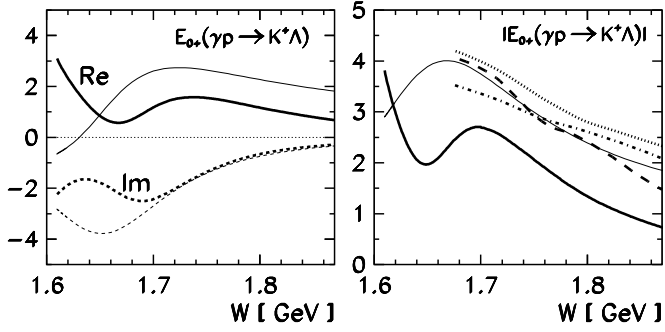}
\end{center}
\caption{The transverse $K^+$-production amplitude $E_{0+}$.
Left panel: real and imaginary parts; our result (thick lines);
KaonMAID \cite{KaonMAID} (thin lines).
Right panel: absolute value of the amplitude;
our result (thick solid line); KaonMAID (thin solid line);
SAID \cite{KaonSAID} (dotted line); 
Bonn-Gatchina \cite{Anisovich10} (dot-dashed line),
JSLT \cite{JuliaDiaz06} (long-dashed line).}
\label{fig:E0Kpro}
\end{figure}

The situation is less clear in the $K\Lambda$ channel displayed 
in fig.~\ref{fig:E0Kpro} since different analyses yield 
apparently conflicting 
results, in particular for the $E_{0+}$ amplitude which
is, however, determined only up to a phase.
Our prediction for the absolute value is consistent with the 
recent compilation of various analyses by Sandorfi and
coauthors \cite{Sandorfi10}.
The real and the imaginary parts of the $E_{0+}$ amplitude
agree with the analysis of the MAID group except in the region
of the $N(1650)$ resonance; the discrepancy can probably be 
attributed to the higher position of the pole of the $K$ matrix 
which is at $W=1690$~MeV in our calculation.
While the results for the inelastic scattering amplitudes
(table~\ref{bg:tableS11} and fig.~\ref{fig:S11ch}) indicate that 
the strength of the $K\Lambda$ coupling is overestimated in our 
model, the results for photoproduction (fig.~\ref{fig:E0Kpro})  
show that our value for this coupling is not inconsistent with 
the phenomenological analyses.

\section{\label{conclusions}Conclusions}

Using our method based on the coupled-channel formalism 
incorporating quark-model quasi-bound states, and the
Cloudy Bag Model to describe the quark states and the
coupling to the mesons, we have been able to reproduce
the main features of $\pi$- and $\gamma$-induced production
of pions, $\eta$ mesons, and kaons.
We have used the same model parameters as in the case of 
the positive-parity resonances, adding only the mixing 
parameter between the two bare-quark states corresponding 
to $N(1535)$ and $N(1650)$, and their bare masses. 
We have not attempted to add further adjustable parameters 
in order to obtain a perfect fit to available data, being 
aware of the limitations of the model which, in general, 
is able to reproduce different observables pertinent to the 
ground state and the excited states within $\sim20$~\%.

The role of the meson cloud turns out to be important in 
two aspects: it enhances the bare baryon-meson coupling and 
improves the behaviour of the helicity amplitudes in the 
region of low $Q^2$.
The enhancement is however not so strong as in the case of 
the positive-parity resonances and the electroexcitation
amplitudes are dominated by the quark contribution.
The enhancement does not lead to a formation of meson-baryon 
quasi-bound states as predicted by some models based on 
the Weinberg-Tomo\-zawa type of meson-baryon interaction.
The critical parameter to distinguish between the two 
mechanisms of resonance formation can be related to the 
effective momentum cut-off which in our case corresponds to 
a rather low value of $\sim 500$~MeV/$c$.

The couplings of the resonances to different inelastic 
channels are reasonably well reproduced particularly in 
the $\eta N$ channel.
At present, the experimental data are insufficient to 
support a reliable multipole analysis of the $K\Lambda$ 
photoproduction channel \cite{KaonSAID}, and there is a 
large spread of predictions for the $E_{0+}$ multipole.  
Our calculation of this channel does not appear to be in 
conflict with any of them.  
We therefore conclude that our assumption about the form of
the quark interaction with the pseudoscalar octet mesons
is sensible.
The $d$-wave pion coupling to $\Delta(1232)$ turns out to
be overestimated; for $\rho N$ channels the available data 
do not allow us to draw more definitive conclusions.

\begin{acknowledgement}
The authors would like to express their thanks to M. D\"oring
for a helpful discussion regarding the $K\Lambda$ channel,
and to I.~Strakovsky for kindly providing us with the latest
set of SAID partial-wave amplitudes.
One of the authors (B. G.) would like to acknowledge the
hospitality he enjoyed during a visit to the University of 
Coimbra and stimulating discussions with M. Fiolhais, 
L. Alvarez-Ruso and P. Alberto.
\end{acknowledgement}


\appendix
\section{\label{vertices}The Cloudy Bag Model meson-quark vertices}

The vertices $V_{lmt}(k)$ in (\ref{Hpi}) are evaluated in the 
Cloudy Bag Model assuming that one of the three quarks is excited 
from the $1s$ state to the $1p_j$ state with the total angular 
momentum $j$ either $1/2$ or $3/2$.
The relevant quark bispinors in the $j\,m_j$ basis are
\begin{eqnarray}
\psi_s(\vec{r}) 
&=& 
   {N_s\over\sqrt{4\pi}\,j_0(\omega_s)}
\left(
  {-\i\,j_0(\omega_s r/R) 
\atop 
  \vec{\sigma}\cdot\hat{\vec{r}}\,j_1(\omega_s r/R)}
\right)\,
  \chi_{m_j}\,,
\nonumber\\
\psi_{p_{1/2}}(\vec{r}) 
&=& 
  {N_{p_{1/2}}\over\sqrt{4\pi}\,j_0(\omega_{p_{1/2}})}
\left(
  {\i\,j_1(\omega_{p_{1/2}}r/R)\vec{\sigma}\cdot\hat{\vec{r}}
\atop
  j_0(\omega_{p_{1/2}}r/R)}
\right)\,
  \chi_{m_j}\,,
\nonumber\\
\psi_{p_{3/2}}(\vec{r}) 
&=& 
   {N_{p_{3/2}}\over\sqrt{6\pi}\,j_1(\omega_{p_{3/2}})}
\left(
    {-\i\,j_1(\omega_{p_{3/2}}r/R) 
\atop 
    \vec{\sigma}\cdot\hat{\vec{r}}\,j_2(\omega_{p_{3/2}}r/R)}
\right)\,
\nonumber\\ && \times
  \sum_{m_sm}\chi_{m_s}\hat{r}_m \CG{\h}{m_s}{1}{m}{\th}{m_j}\,.
\nonumber
\end{eqnarray}
Here $\chi_m$ is the spinor for spin $\half$, $R$ is the bag radius, 
$\omega_s=2.043$, $\omega_{p_{1/2}}=3.811$, $\omega_{p_{3/2}}=3.204$, and
$$
   N_s^2      = {\omega_s\over2R^3(\omega_s-1)}\,,
\qquad
   N_{p_{1/2}}^2 = {\omega_{p_{1/2}}\over2R^3(\omega_{p_{1/2}}+1)}\,,
$$
$$
   N_{p_{3/2}}^2 = {9\omega_{p_{3/2}}\over4R^3(\omega_{p_{3/2}}-2)}\,.
$$

For the {\em quark pion coupling\/} we obtain
\begin{eqnarray}
V^\pi_{l=0,t}(k) &=& {1\over2f_\pi}\sqrt{\omega_{p_{1/2}}\omega_s\over 
    (\omega_{p_{1/2}}+1)(\omega_s-1)}\,
     {1\over2\pi}\,{k^2\over\sqrt{\omega_k}}\,{j_0(kR)\over kR}
\nonumber\\ && \times
     \sum_{i=1}^3 \tau_t(i)\, 
\mathcal{P}_{sp}(i)\,,
\nonumber\\
V^\pi_{1mt}(k) &=& {1\over2f_\pi}{\omega_s\over(\omega_s-1)}\,
     {1\over2\pi}\,{1\over\sqrt3}\,
     {k^2\over\sqrt{\omega_k}}\,{j_1(kR)\over kR}
     \sum_{i=1}^3 \tau_t(i)
\nonumber\\
&&\times\left(\sigma_{m}(i)
+ r_{p_{1/2}} S^{[\h]}_{1m}(i)
+ r_{p_{3/2}} S^{[\th]}_{1m}(i)\right)\,,
\nonumber\\
V^\pi_{2mt}(k) &=& {1\over2f_\pi}\sqrt{\omega_{p_{3/2}}\omega_s\over
     (\omega_{p_{3/2}}-2)(\omega_s-1)}\,
     {\sqrt2\over2\pi}\,{k^2\over\sqrt{\omega_k}}\,{j_2(kR)\over kR}
\nonumber\\ && \times
     \sum_{i=1}^3 \tau_t(i)\,\Sigma^{[\h\th]}_{2m}(i)\,.
\nonumber
\end{eqnarray}
Here 
\begin{eqnarray}
\mathcal{P}_{sp}&=&\sum_{m_j}|sm_j\rangle\langle p_{1/2}m_j|\,,
\nonumber\\
S^{[\h]}_{1m}
&=&  \sqrt3\kern-3pt\sum_{m_jm_j'}\kern-3pt\CG{\h}{m_j'}{1}{m}{\h}{m_j}
   |p_{1/2}m_j\rangle\langle p_{1/2}m_j'|\,,
\nonumber\\
\Sigma^{[\h\th]}_{2m}
    &=& \kern-3pt\sum_{m_sm_j}\kern-3pt\CG{\th}{m_j}{2}{m}{\h}{m_s}
      |sm_s\rangle\langle p_{3/2}m_j|\,,
\nonumber\\
S^{[\th]}_{1m}
 &=&{\textstyle {\sqrt{15}\over2}}\kern-3pt\sum_{m_jm_j'}\kern-3pt
  \CG{\th}{m_j'}{1}{m}{\th}{m_j}
  |p_{3/2}m_j\rangle\langle p_{3/2}m_j'|\,,
\end{eqnarray}
and
$$
r_{p_{1/2}} = {\omega_{p_{1/2}}(\omega_s-1)\over \omega_s(\omega_{p_{1/2}}+1)}\,,
\qquad
r_{p_{3/2}} = {2\omega_{p_{3/2}}(\omega_s-1)\over 5\omega_s(\omega_{p_{3/2}}-2)}\,.
$$

For the {\em $s$-wave\/}
octet $\eta$ and {\em $K$ mesons\/} we assume 
the flavour SU(3) symmetry yielding
\begin{eqnarray}
   V^\eta(k) &=& {1\over2f_\pi}\sqrt{\omega_{p_{1/2}}\omega_s
   \over(\omega_{p_{1/2}}+1)(\omega_s-1)}\,
     {1\over2\pi}\,{k^2\over\sqrt{\omega_k}}\,{j_0(kR)\over kR}
\nonumber\\ && \times
    \sum_{i=1}^3
          \lambda_8(i)\,\mathcal{P}_{sp}(i)\,,
\nonumber\\
   V^K_{t}(k) &=& {1\over2f_K}\sqrt{\omega_{p_{1/2}}\omega_s
\over(\omega_{p_{1/2}}+1)(\omega_s-1)}\,
     {1\over2\pi}\,{k^2\over\sqrt{\omega_k}}\,{j_0(kR)\over kR}
\nonumber\\ && \times
\sum_{i=1}^3
          (V_{t}(i)+U_{t}(i))\,\mathcal{P}_{sp}(i)\,,
\nonumber
\end{eqnarray}
with $t=\pm\half$, $V_{\pm t}= (\lambda_4\pm\i\lambda_5)/\sqrt2$
$U_{\pm t} = (\lambda_6\pm\i\lambda_7)/\sqrt2$,
$f_\eta=1.20\,f_\pi$~\cite{PDG06}, and $f_K=1.20\,f_\pi$~\cite{PDG}.

For the coupling of negative parity states to $\rho N$, the 
dominant contribution is expected to arise from the transverse 
$\rho$-mesons with the total $J=1$ and the orbital angular momentum 
of the $\rho N$ system equal to either 0 or 2. 
Assuming that the $\rho$ meson couples to the quarks only on the bag 
surface \cite{Thomas83} in the form $\gamma^\mu\rho_\mu$,
we find (note that $m$ in (\ref{Hpi}) and below refers to 
the total angular momentum rather than to the orbital one):
\begin{eqnarray}
V^\rho_{l=0 mt}(k)&=&
{1\over2f_\rho}\sqrt{\omega_s\over(\omega_s-1)}\,
     {1\over2\pi}\,
     {k^2\over\sqrt{\omega_k}}\,{j_0(kR)\over kR}\sum_{i}\tau_t(i)
\nonumber\\
&&\kern-30pt\times
    \left({\sqrt8\over3}\,\sqrt{\omega_{p_{1/2}}
\over \omega_{p_{1/2}}+1}\,\Sigma_{1m}^{[\h]}
    + 3\sqrt{\omega_{p_{3/2}}\over \omega_{p_{3/2}}-2}\, 
      \Sigma_{1m}^{[\h\th]}(i)\right)\,,
\nonumber\\
V^\rho_{l=2 mt}(k)&=&
{1\over2f_\rho}\sqrt{\omega_{p_{3/2}}\omega_s
\over(\omega_{p_{3/2}}-2)(\omega_s-1)}\,
     {1\over2\pi}\,{1\over3}\,
     {k^2\over\sqrt{\omega_k}}\,{j_2(kR)\over kR}
\nonumber\\ && \times
\sum_{i=1}^3
     \tau_t(i)\Sigma_{1m}^{[\h\th]}(i)\,.
\nonumber
\end{eqnarray}
Here 
\begin{eqnarray}
\Sigma_{1m}^{[\h]}
    &=& \kern-3pt\sum_{m_sm_j}\kern-3pt\CG{\h}{m_j}{1}{m}{\h}{m_s}
      |sm_s\rangle\langle p_{1/2}m_j|\,,
\nonumber\\
\Sigma_{1m}^{[\h\th]}
    &=& \kern-3pt\sum_{m_sm_j}\kern-3pt\CG{\th}{m_j}{1}{m}{\h}{m_s}
      |sm_s\rangle\langle p_{3/2}m_j|\,,
\end{eqnarray}
and
$f_\rho$ is the $\rho$-meson decay constant with the
experimental value $208$~MeV.

The peculiar oscillating shape of the CBM form factor has 
little influence in the case of the $p$ and $d$-wave pions but 
leads to the unphysical behaviour of the $s$-wave scattering
amplitude since it crosses zero already at $W \sim 1950$~MeV.
We have cured this problem by replacing  $j_0(kR)$ by an 
exponential tail for $k>1.6/R$ in such a way as not to alter 
the value of the self energy integral.

\end{document}